\documentclass[12pt]{article}
\def\nabstar#1{\nabla\kern-0.5pt\smash{\raise 4.5pt\hbox{$\ast$}}
               \kern-4.5pt_{#1}}

\def\drvstar#1{\partial\kern-0.5pt\smash{\raise 4.5pt\hbox{$\ast$}}
               \kern-5.0pt_{#1}}

\def\newline{\relax\ifhmode\null\hfil\break\else\nonhmodeerr@\newline\fi}
\def\frac#1#2{{#1\over#2}}
\def\text#1{{\hbox{\rm #1}}}
\def\flushpar{{\par \noindent}}

\newcommand{\beq}{\begin{equation}}
\newcommand{\eeq}{\end{equation}}
\newcommand{\bea}{\begin{eqnarray}}
\newcommand{\eea}{\end{eqnarray}}
\def\Id{ \mbox{1\hspace{-1.2mm}I} }
\def\BE{\begin{equation}}
\def\EE{\end{equation}}
\def\BA{\begin{eqnarray}}
\def\EA{\end{eqnarray}}
\def\BAN{\begin{eqnarray*}}
\def\EAN{\end{eqnarray*}}
\def\LL{\left}
\def\RR{\right}

\def\nn{\nonumber\\}

\def\det{\mbox{det}}

\def\gm5{\gamma_5}

\input epsf.sty
\newdimen\psfigsize
\def\psfigure#1 #2 #3 #4 #5{
    \begin{figure}[tbh]
      \begin{center}
      \vbox{
        \null\vskip-0.2in\hskip#2
        \epsfxsize=#1
        \epsfbox{#4}
        \vskip -0.3in
        \caption {#5 \label{#3}}
        \vskip 0.0 true in plus 0.3 true in
      }
      \end{center}
   \end{figure}
}
\usepackage{graphicx}
\begin{document}
\thispagestyle{empty}
\begin{flushright}
NTUTH-03-505D \\
June 2003
\end{flushright}
\bigskip\bigskip\bigskip
\vskip 0.5truecm
\begin{center}
{\LARGE {
A note on Neuberger's double pass algorithm}}
\end{center}
\vskip 2.0truecm
\centerline{Ting-Wai Chiu and Tung-Han Hsieh}
\vskip 0.5cm
\centerline{Department of Physics, National Taiwan University}
\centerline{Taipei, Taiwan 106, Taiwan.}
\vskip 1.5truecm
\bigskip \nopagebreak \begin{abstract}
\noindent

We analyze Neuberger's double pass algorithm for
the matrix-vector multiplication $ R(H) \cdot Y $
(where $ R(H) $ is $ (n-1,n) $-th degree rational polynomial of
positive definite operator $H$), and show that the number
of floating point operations is independent of the degree $ n $,
provided that the number of sites is much larger
than the number of iterations in the conjugate gradient.
This implies that the matrix-vector product 
$ (H)^{-1/2} Y \simeq R^{(n-1,n)}(H) \cdot Y $
can be approximated to very high precision with sufficiently
large $ n $, without noticeably extra costs.
Further, we show that there exists a threshold $ n_T $
such that the double pass is faster than the single pass
for $ n > n_T $, where $ n_T \simeq 12 - 25 $ for most platforms.



\end{abstract}
\vskip 1.5cm

\newpage\setcounter{page}1

\section{Introduction}

In 1998, Neuberger proposed the nested conjugate gradient 
\cite{Neuberger:1998my} for solving the propagator of the
overlap-Dirac operator \cite{Neuberger:1998fp}
\bea
\label{eq:D}
D = m_0 \left( 1+\gamma_5 \frac{H_w}{\sqrt{H_w^2}} \right) \ ,
\eea
with the sign function $ \mbox{sgn}(H_w) \equiv H_w (H_w^2)^{-1/2} $ 
approximated by the polar approximation
\bea
\label{eq:S_polar}
S(H_w) = \frac{H_w}{n} \sum_{l=1}^{n} \frac{b_l}{H_w^2 + d_l}
         \equiv H_w R^{(n-1,n)}(H_w^2) \ ,  
\eea
where $ H_w = \gamma_5 D_w $, and $ D_w $ is the standard Wilson-Dirac 
operator plus a negative parameter $ -m_0 $ ($ 0 < m_0 < 2 $), 
and the coefficients $ b_l $ and $ d_l $ are:
\BAN
\label{eq:polar_bd}
b_l = \sec^2 \left[ \frac{\pi}{2n} \left( l-\frac{1}{2} \right) \right] \ ,
\hspace{4mm}
d_l = \tan^2 \left[ \frac{\pi}{2n} \left( l-\frac{1}{2} \right) \right] \ .
\EAN 

In principle, any column vector of
$ D^{-1} = D^{\dagger} ( D D^{\dagger} )^{-1} $ can be
obtained by solving the system  
\bea
\label{eq:outer_CG}
D D^{\dagger} Y  
= m_0^2 [ 2 + \gamma_5 S(H_w)+ S(H_w) \gamma_5 ] Y = \Id
\eea
with conjugate gradient, provided that the matrix-vector 
product $ S(H_w) Y $ can be carried out. Writing
\bea
S(H_w) Y = \frac{H_w}{n} \sum_{l=1}^{n} b_l Z^{(l)} \ ,
\eea
one can obtain $\{ Z^{(l)} \}$ by solving the system
\bea
\label{eq:inner_CG}
(H_w^2 + d_l ) Z^{(l)} = Y
\eea 
with multi-shift conjugate gradient \cite{Frommer:1995ik,Jegerlehner:1996pm}.
In other words, each (outer) CG iteration in (\ref{eq:outer_CG}) contains 
a complete (inner) CG loop (\ref{eq:inner_CG}), 
i.e., nested conjugate gradient.

Evidently, the overhead for the nested conjugate gradient is the 
execution time for the inner conjugate gradient loop (\ref{eq:inner_CG})
as well as the memory space it requires, i.e., $ (2n + 3)$ large vectors, 
each of $ 12 N_{site} $ double complex numbers, where $ N_{site} $ 
is the number of sites, and 12 = 3 (color) $\times$ 4 (Dirac) 
is the degrees of freedom at each site for QCD.
The memory storage becomes prohibitive for large lattices since $ n $ is
often required to be larger than 16 in order to achieve a reliable
approximation for the sign function. To minimize the memory storage for
$\{ P^{(l)}, Z^{(l)} \}$, Neuberger \cite{Neuberger:1998jk} observed that
one only needs the linear combination $ \sum_{l=1}^{n} b_l Z^{(l)} $
rather than each $ Z^{(l)} $ individually. Since $ Z_i^{(l)} $ and
its conjugate vector $ P_i^{(l)} $ at the $i$-th iteration of the
inner CG are linear combinations of their preceedents  
$ \{ P_j^{(l)},  Z_{j}^{(l)}, j=0, \cdots, i-1 \} $ in the
iteration process, thus it is possible to obtain their updating 
coefficients $ \{ \alpha_i^{(l)}, \beta_i^{(l)} \} $ in the first pass,
and then use them to update the sum 
$ \sum_{l=1}^{n} b_l Z^{(l)} $
successively in the second pass, with memory storage of only 5 vectors, 
independent of the degree $ n $ of the rational polynomial $ R^{(n-1,n)} $.

At first sight, the double pass algorithm seems to be slower than the
single pass algorithm. However, in the test run
(with $ SU(2) $ gauge field on the $ 8^3 $ lattice), 
Neuberger found that the double pass actually ran faster by $30\%$ than 
the single pass, and remarked that the speedup most likely reflects 
the cache usage in the testing platform, the SGI O2000 
(with four processors, each with 4MB cache memory). 

In this paper, we analyze the number of floating point operations
($F_2$) for the double pass algorithm, 
and show that it is independent of the degree $ n $ of $ R^{(n-1,n)} $,
provided that the number of lattice sites ($ N_{site} $)
is much larger than the number of iterations ($ L_{i} $) of the
CG loop. The last condition is amply satisfied even for a small lattice
(e.g., $ N_{site} = 8^3 \times 24 = 12288 $), since $ L_i $ is usually less
than 1000 (after the low-lying eigenmodes of $ H_w^2 $ are projected out).
On the other hand, the number of floating point operations ($F_1$) for
the single pass is a linearly increasing function of $ n $. It follows
that there exists a threshold $ n_F $ such that $ F_2 \le F_1 $ for
$ n \ge n_F $, where the value of $ n_F $ depends on the
implementation of the algorithms ($n_F \simeq 59$ for our codes).
Corresponding to the number of floating point operations, we also obtain 
the formulas for the CPU times ($ T_1 $ and $ T_2 $) for the single and 
the double pass algorithms. Further, we show that there exists a 
threshold $ n_T $ such that the double pass is faster than the 
single pass\footnote{
In this paper, we only consider the (faster) single pass algorithm 
in which the vectors $ P^{(l)} $ and $ Z^{(l)} $ ($l=2,\cdots,n$) 
are not updated after $ Z^{(l)} $ converges.}
for $ n > n_T $, where $ n_T \simeq 12 - 25 $ (for most platforms), 
which is quite smaller than the threshold $ n_F \simeq 59 $ for
the number of floating point operations. 
By timing the speed of each subroutine, we can account for  
the extra slow down in the single pass algorithm, which is
unlikely to be eliminated, due to the memory bandwidth, 
a generic weakness of any computational system. 
Thus, in general (for most vector or superscalar machines), 
one may find that the double pass is faster than the single pass,
for $ n > n_T \simeq 12 - 25 $. This explains why in Neuberger's
test run, even at $ n = 32 $, the double pass is already
faster than the single pass by 30\%. In fact, we find that 
DEC alpha XP1000 and IBM SP2 SMP also have $ 30\% $ speedup
at $ n=32 $ (see Table \ref{tab:P4_IBM_DEC}), which agrees
with the theoretical estimate (\ref{eq:speedup}) 
using the CPU time formulas (\ref{eq:T1a})-(\ref{eq:T2a}). 

Nevertheless, the most interesting result is that
{\it the speed of the double pass algorithm is almost
independent of the degree $ n $}.
This implies that {\it the matrix-vector product 
$ (H_w^2)^{-1/2} Y \simeq R^{(n-1,n)}(H_w^2) Y $
can be approximated to very high precision with sufficiently
large $ n $, without noticeably extra costs.}

The outline of this paper is as follows.
In section 2, we outline the single and double pass
algorithms for the iteration of the (inner) CG loop
(\ref{eq:inner_CG}), and analyze their major differences.
In section 3, we estimate the number of (double precision)
floating point operations as well as the CPU time
for the single and double pass algorithms respectively,
and show that there exists a threshold $ n_T $
such that the double pass is faster than the single pass
for $ n > n_T $. In section 4, we perform some tests.
In section 5, we conclude with some remarks.

\section{The single and double pass algorithms}

In the section, we outline the single and the double pass
algorithms for the inner CG loop (\ref{eq:inner_CG}),
and point out their major differences.

For the single pass algorithm, with the input vector $ Y $,
we initialize the vector variables
$\{Z^{(l)}, P^{(l)} \}, R, A, B $ and the
scalar variables $ \alpha, \beta, \{ \gamma^{(l)} \}$ as
\BAN
&& Z_{0}^{(l)}=0, \ P_{0}^{(l)}=Y,  \hspace{4mm} l=1,\cdots, n    \nn
&& R_0=Y, \nn
&& \alpha_{-1} = 1,       \nn
&& \beta_{0} = 0,     \nn
&& \gamma_{-1}^{(l)}=\gamma_{0}^{(l)}=1, \hspace{4mm} l=1,\cdots, n
\EAN
Then we iterate ($ j=0,1, \cdots $) according to:
\bea
\label{eq:A}
A_j &=& H_w P_j^{(1)}             \\
\label{eq:B}
B_j &=& H_w A_j + d_1 P_j^{(1)} = (H_w^2+d_1) P_j^{(1)} \\
\label{eq:alpha}
\alpha_j &=& {\LL<R_j| R_j\RR> \over \LL<P_j^{(1)}|B_j\RR>} \\
\label{eq:R}
R_{j+1}       &=& R_{j} - \alpha_j B_j \\
\label{eq:beta}
\beta_{j+1}   &=& \LL<R_{j+1}|R_{j+1}\RR> \over \LL<R_{j}|R_{j}\RR> \\
\label{eq:P1}
P_{j+1}^{(1)} &=& R_{j+1} + \beta_{j+1} P_j^{(1)} \\
\label{eq:Z1}
Z_{j+1}^{(1)} &=& Z_j^{(1)} + \alpha_j P_j^{(1)} 
\eea
together with the following updates for $l=2,\cdots,n$:
\bea
\label{eq:gamma}
\gamma_{j+1}^{(l)} &=&
\frac{ \gamma_j^{(l)} \gamma_{j-1}^{(l)} \alpha_{j-1} }
     { \alpha_j \beta_j ( \gamma_{j-1}^{(l)}-\gamma_j^{(l)} ) +
       \gamma_{j-1}^{(l)} \alpha_{j-1} (1+\alpha_j(d_l-d_1)) } \\
\label{eq:Pl} 
P_{j+1}^{(l)} &=& \gamma_{j+1}^{(l)} R_{j+1} + \beta_{j+1} 
\left(\frac{\gamma_{j+1}^{(l)}}{\gamma_j^{(l)}}\right)^2 P_j^{(l)} \\
\label{eq:Zl}
Z_{j+1}^{(l)} &=& Z_j^{(l)} +
    \alpha_j \frac{ \gamma_{j+1}^{(l)} }{ \gamma_j^{(l)} } P_{j}^{(l)} 
          \ . 
\eea
The loop terminates at the $i$-th iteration if
$\sqrt{\LL<R_{i+1}|R_{i+1}\RR>/\LL<Y|Y\RR>} $ is less than
the tolerance (tol).

Since we are only interested in the linear combination 
$ \sum_{l=1}^{n} b_l Z_{i+1}^{(l)} $, in which each $ Z_{i+1}^{(l)} $
can be expressed in terms of $ \{ R_j, j=0,\cdots,i \} $, thus
we can write  
\bea
\label{eq:sum}
\sum_{l=1}^{n} b_l Z_{i+1}^{(l)} = \sum_{j=0}^{i} c_j R_j  
\eea
where $ c_j $ can be derived as \cite{Neuberger:1998jk}
\bea
\label{eq:cij}
c_j = \sum^{i-j}_{m=0} \left[ \alpha_{j+m} \ \delta_m
          \left( b_1 + \sum_{l=2}^{n} b_{l}
          \frac{\gamma^{(l)}_{m+j+1}\gamma^{(l)}_{m+j}}{\gamma^{(l)}_{j}}
\right) \right] \ ,  
\eea
with
\bea
\delta_m = \LL\{ \begin{array}{ll}
           \prod^m_{k=1} \beta_{j+k} \ , &  \mbox{for } m > 0, \\
           1 \ , &  \mbox{for } m = 0. \end{array} \RR.
\eea
Therefore, the r.h.s. of (\ref{eq:sum}) can be evaluated with 
the CG loop (\ref{eq:A})-(\ref{eq:P1}), requiring only the 
storage of 5 large vectors 
($ A, B, R, P^{(1)}, T=\sum c_j R_j $), 
provided that the coefficients $ \{ c_j, j=0,\cdots,i \} $
are known. However, from (\ref{eq:cij}), the determination of
$ c_j $ at any $j$-th iteration requires some values of
$ \{ \alpha \} $, $ \{ \beta \} $ and $ \{ \gamma \} $ 
which can only be obtained in later iterations.
Thus we have to run the first pass, i.e.,
the CG loop (\ref{eq:A})-(\ref{eq:P1}),
to obtain all coefficients of $ \{ \alpha \} $, $ \{ \beta \} $ 
up to the convergence point $i$, and then compute all
$ \{ c_j, j=0,\cdots,i \} $ according to (\ref{eq:cij}) and (\ref{eq:gamma}).
Finally, we run the second pass,
i.e., going through (\ref{eq:A}), (\ref{eq:B}), (\ref{eq:R}),  
(\ref{eq:P1}), and adding $ c_j R_j $ to the r.h.s. of (\ref{eq:sum}), 
successively from $ j=0 $ to the convergence point $i$.

Evidently, all operations in (\ref{eq:A})-(\ref{eq:Z1}) and
(\ref{eq:Pl})-(\ref{eq:Zl}) are proportional to the number of 
lattice sites $ N_{site} $ times the number of iterations $ L_i $.
On the other hand, the computations of the coefficients 
$\{ \gamma \}$ (\ref{eq:gamma}) and $ \{ c_j \} $ (\ref{eq:cij})
do not depend on $ N_{site} $, but only on $ L_{i} $ (up to 
a small term proportional to $L_i^3$).   
Thus, for $ N_{site} \gg L_i $, we can neglect the computation of 
$ \{ c_j \} $ (\ref{eq:cij}), and focus on the major difference between 
the single pass and the double pass, namely, the number of operations 
in (\ref{eq:Pl})-(\ref{eq:Zl}), which is proportional to 
$ (n-1) N_{site} L_i $, versus the number of operations 
in (\ref{eq:A}), (\ref{eq:B}), (\ref{eq:R}), (\ref{eq:P1})
plus the vector update on the r.h.s. of (\ref{eq:sum}), which is
proportional to $ N_{site} L_i $. 
Obviously, {\it the number of floating-point operations in the single pass 
is a linearly increasing function of $ n $, while that of the double pass 
is independent of $ n $, thus it follows that the double pass must be faster 
than the single pass for sufficiently large $ n $.}

In the next section, we estimate the number of floating point
operations as well as the CPU time, for the single pass and the double pass
respectively. Even though our countings are based on our codes,
they serve to illustrate the general
features of the single and the double pass algorithms, which 
are valid for any software implementations and/or machines.

\section{The CPU time and the number of floating point operations}

For our codes, the number of floating point operations
for the single pass is 
\bea
\label{eq:F1}
F_1 = N_{site} L_{i} [3552+120(n-1)p]
                    + N_{site} (288 n_{ev} + 48n + 1776) \ ,   
\eea
while for the double pass is 
\bea
\label{eq:F2}
F_2 &=&  6888 \ N_{site} L_{i}  
        + N_{site} (288 n_{ev}-1656)
        \nn
    & & + \left[\frac{L_{i}^3}{6} + L_{i}^2 (2n-1)
        + L_{i} \left(13 n -\frac{73}{6} \right) - 7 n + 7 \right] q
\eea
where $ N_{site} $ is the number of sites of the lattice,
$ L_{i} $ is the number of iterations of the CG loop, 
$ n_{ev} $ is the number of projected eigenmodes of $ H_w^2 $,
and $ n $ is the degree of the rational polynomial $ R^{(n-1,n)} $.
In the single pass, Eq. (\ref{eq:F1}), $ (n-1)p $ is the effective
number of the $ (n-1) $ updates in (\ref{eq:Pl})-(\ref{eq:Zl}), since
$ P^{(l)} $ and $ Z^{(l)} $ are not updated after $ Z^{(l)} $ converges.
The value of $ p $ depends on the convergence criteria as well as
the rational polynomial $ R^{(n-1,n)} $ and its argument.
Similarly, in the double pass, the sum in (\ref{eq:cij}) only
includes the terms which have not yet converged at the iteration $ j $,
and the reduction in the number of floating point operations
can be taken into account by the factor $ q $ in (\ref{eq:F2}).
(The value of $ q $ is about 0.95 for convergence up to the 
zero in the IEEE double precision representation). 

Taking into account of different speeds of various floating point
operations, we estimate the CPU time for the single pass and the
double pass as follows.
\bea
\label{eq:T1}
T_1 & = &
  N_{site} L_i [ 192 t_b + 72 t_c + 3288 t_d +
                     (48 t_b+72 t_a)(n-1)p ] \nn
  & & +N_{site} (288 n_{ev} t_e + 48 n t_b + 24 t_a + 108 t_c + 1644 t_d) \\
\label{eq:T2}
T_2 & = & N_{site} L_i ( 240t_b+72t_c+6576t_d ) \nn
    & & +N_{site} ( 288 n_{ev} t_e + 24 t_a -144 t_b+108 t_c-1644 t_d ) \nn
    & &
    + q \left[\frac{L_{i}^3}{6} + L_{i}^2 (2n-1)
        + L_{i} \left(13 n -\frac{73}{6} \right) - 7 n + 7 \right] t_f
\eea
where $ t_a, t_b, t_c $ and $ t_d $ denote the average CPU time per
floating-point operation (FPO) for the four different kinds of vector 
operations (a)-(d) listed in Table \ref{tab:time} respectively, 
$ t_e $ the average CPU time per FPO for constructing the complementary 
vector from the projected eigenmodes of $ H_w^2 $, 
and $ t_f $ the time for computing the coefficients (\ref{eq:cij}) in 
the double pass.
Note that setting $ t_a = t_b = t_c = t_d = t_e = t_f = 1 $ in
(\ref{eq:T1}) and (\ref{eq:T2}) reproduce (\ref{eq:F1}) and (\ref{eq:F2})
respectively.

It should be emphasized that {\it the numerical values of the
constants and coefficients in (\ref{eq:F1})-(\ref{eq:T2})
may vary slightly from one implementation to another, however,
the number of different terms and their functional dependences 
on the variables ($ N_{site} $, $ L_i $, $ n $, $ n_{ev} $, $ p $, $ q $,
$ t_a, t_b, t_c, t_d, t_e $, and $ t_f $) should be the same for any 
codes of the single and double pass algorithms.}

For the double pass, it is clear that the first term on the r.h.s. 
of Eq. (\ref{eq:F2}) is the most significant part, since
the number of lattice site ($ N_{site} $) is usually much larger
than the number of iterations ($ L_{i} $) of the CG loop
such that the second and the third terms on the r.h.s. of (\ref{eq:F2})
can be neglected. For example, $ N_{site} = 8^3 \times 24 $, 
$ L_i = 1000 $, $ n = 16 $, $ n_{ev} = 32 $ and $ q = 0.95 $, 
then the first term is $ 6888 N_{site} L_{i} \simeq 8.5 \times 10^{10} $, 
while the sum of the second and the third terms only gives 
$ \sim 2.8 \times 10^8 $. Thus we can single out the most significant 
part of $ F_2 $,
\bea
\label{eq:F2a}
F_2 \simeq 6888 \ N_{site} L_{i} \ ,  
\eea
which comes from the first pass (\ref{eq:A})-(\ref{eq:P1}) and
the second pass (\ref{eq:A}), (\ref{eq:B}), (\ref{eq:R}), (\ref{eq:P1})
plus the vector update on the r.h.s. of (\ref{eq:sum}).
Similarly, for the single pass, the most significant part of $F_1$
is the first term on the r.h.s. of (\ref{eq:F1})
\bea
\label{eq:F1a}
F_1 \simeq  N_{site} L_{i} [3552+120(n-1)p] \ , 
\eea
which comes from the operations in (\ref{eq:A})-(\ref{eq:Z1}),
and (\ref{eq:Pl})-(\ref{eq:Zl}). 

Evidently, from (\ref{eq:F1a}) and (\ref{eq:F2a}),
$ F_1 $ is a linearly increasing function of $ n $ 
while $ F_2 $ is independent of $ n $. 
Thus it follows that there exists a threshold $ n_F $
such that $ F_2 < F_1 $ for $ n > n_F $.
From (\ref{eq:F1a}) and (\ref{eq:F2a}), we obtain
the threshold $ n_F $,
\bea
\label{eq:nF}
n_F = 1 + \frac{139}{5 p} \ ,
\eea
where the value of $ p $ depends on the convergence criterion
for removing $ \{ P^{(l)}, Z^{(l)} \} $ from the updating list,  
as well as the rational polynomial $ R^{(n-1,n)} $ and its argument.
For our codes and the tests in the next section, $ p \simeq 0.48 $, 
thus we have  
\bea
\label{eq:nFa}
n_F \simeq 59 \ . 
\eea

Assuming $ N_{site} \gg L_i $, we obtain the most significant parts 
of the CPU times (\ref{eq:T1})-(\ref{eq:T2}) as
\bea
\label{eq:T1a}
T_1 &\simeq&  N_{site} L_i [ 192 t_b + 72 t_c + 3288 t_d +
                            (48 t_b+72 t_a)(n-1)p ]  \ , \\
\label{eq:T2a}
T_2 &\simeq&  N_{site} L_i ( 240 t_b + 72 t_c + 6576 t_d ) \ .
\eea
Obviously, from (\ref{eq:T1a}) and (\ref{eq:T2a}),
there exists a threshold
\bea
\label{eq:nT}
n_T =  1 + \frac{2 t_b + 137 t_d}{(2 t_b + 3 t_a)p}
\eea
such that $ T_2 < T_1 $
(the double pass is faster than the single pass) for $ n > n_T $.

Even though the countings in (\ref{eq:T1}) and (\ref{eq:T2}) are based
on our codes (for $ R^{(n-1,n)} $ with argument $ H_w^2 $), the
essential features of (\ref{eq:T1}) and (\ref{eq:T2}) should be common
to all implementaions of the single and the double pass
algorithms. In other words, the numerical coefficients in (\ref{eq:T1a})
and (\ref{eq:T2a}) may change from one implementaion to another,
however, {\it the existence of a threshold $ n_T $ must hold
for any implementation.}

Now it is interesting to compare $ n_T $ with $ n_F $. 
From (\ref{eq:nF}) and (\ref{eq:nT}), one immediately sees that 
$ n_T < n_F $ if     
\bea
\label{eq:tdab}
19 t_d < 11 t_a + 7 t_b
\eea
is satisfied\footnote{Note that the inequality (\ref{eq:tdab}) is more
restrictive than $ 685 t_d < 417 t_a + 268 t_b $.}.

In practice, it turns out that $ t_a/t_d > 2 $
and $ t_b/t_d > 3 $ for most systems 
(Tables \ref{tab:time}-\ref{tab:time_1}).
Thus, $ n_T \simeq 12 - 25 $, which is quite smaller than $ n_F \simeq 59 $.

The speedup of the double pass with respect to the single pass
(for $ n > n_T $) can be defined as
\bea
S = \frac{T_1 - T_2}{T_2}
\eea
which is estimated to be
\bea
\label{eq:speedup}
S \simeq \frac{(3 t_a + 2 t_b) p}{10 t_b + 3 t_c + 274 t_d} (n-n_T)
\eea
where Eqs. (\ref{eq:T1a})-(\ref{eq:nT}) have been used.

In Table \ref{tab:time}, we list our measurements of $ t_a, t_b, t_c $ 
and $ t_d $ for four different hardware configurations of Pentium 4, 
i.e., two different Rambuses of faster/slower (PC1066/PC800) speed, 
and with/without SSE2 (the vector processing unit of Pentium 4) codes.

Substituting the values of $ t_a $, $ t_b $ and $ t_d $
into (\ref{eq:nT}),  
we obtain the theoretical estimates for the threshold $ n_T $,
\bea
\label{eq:nTa}
n_T \simeq \LL\{ \begin{array}{ll}
            12, &  \mbox{Pentium 4, PC800, with SSE2},  \\
            22, &  \mbox{Pentium 4, PC800}, \\
            13, &  \mbox{Pentium 4, PC1066, with SSE2}, \\
            25, &  \mbox{Pentium 4, PC1066}, \end{array} \RR.
\eea
where $ p \simeq 0.48 $ has been used. 

\begin{table}
\caption{The average CPU time (in unit of nano-second) per
 floating point operation (FPO) for 4 different kinds of
 matrix-vector operations in the single and double pass algorithms.
 The CPU is Pentium 4 (2.53 GHz), with one Gbyte Rambus (PC800 or PC1066).}
\begin{center}
\begin{tabular}{c|c|c|c|c}
& Operation & \# of FPO & \multicolumn{2}{c}{CPU time(ns) per FPO}
  \\
& &  & \multicolumn{1}{c}{PC800} & \multicolumn{1}{c}{PC1066} \\
& &  & SSE2 on \hspace{2mm}  off & SSE2 on \hspace{2mm}  off   \\
\hline
\hline
(a) & $ |A\rangle = c_1|A\rangle + c_2 |B\rangle $
    & $ 72 \times N_{site} $
    & 3.720 \hspace{4mm} 3.721  & 2.977 \hspace{4mm} 3.016    \\
(b) & $ |V\rangle = |A\rangle + c |B\rangle $
    & $ 48 \times N_{site} $
    & 5.521 \hspace{4mm} 5.522  & 4.330 \hspace{4mm} 4.429   \\
(c) & $ \alpha = \langle V|V\rangle $
    & $ 36 \times N_{site}$
    & 4.249 \hspace{4mm} 4.251  & 3.312 \hspace{4mm} 3.340   \\
(d) & $ |A\rangle = H_w | B \rangle $
    & $ 1644 \times N_{site} $
    & 0.764 \hspace{4mm} 1.535  & 0.686 \hspace{4mm} 1.440    \\
\end{tabular}
\end{center}
\label{tab:time}
\end{table}

Note that for each hardware configuration in Table \ref{tab:time},
the average CPU time per FPO of the simple vector operations 
(a)-(c) is much longer than that of (d), Wilson matrix times vector.
A simplified explanation\footnote{ 
It should be emphasized that the mechansim of the interactions
between the CPU and the RAM is a rather complicated process,
which is beyond the scope of this paper.} 
is as follows. Since all these four vector operations
involve long vectors, the CPU and its cache cannot hold all data at once.
Thus it is necessary to transfer the data from/to the memory
successively, every time after the CPU completes its operations 
on a portion of the vectors. However, for any system, 
the memory bandwidth is limited. Thus, there is a time interval
between consecutive sets of data transferring to/from the CPU.
Therefore, if the CPU finishes a computation before the next
set of data is ready, then it would waste its cycles in idling. 
Since any one of the vector operations (a)-(c) is rather simple,
the CPU finishes a computation at a speed faster than that of
transferring data from/to the memory, thus the CPU ends up wasting a 
significant fraction of time in idling.
On the other hand, for the vector operation (d),
the number of FPO is much more than that of any one of (a)-(c),
thus when the CPU completes its operations on a portion of the vectors,
the next set of data might have been ready, so the CPU does not waste
much time in the memory I/O.
This explains why the average CPU time per FPO of (a)-(c)
is much longer than that of (d).
Further, this simple picture also explains why turning on SSE2 of Pentium 4 
(see Table \ref{tab:time}) doubles the speed of (d) but has 
no speedups for (a)-(c), since the bottleneck of (a)-(c) is essentially
due to the memory bandwidth rather than the speed of the CPU.
 
If the memory bandwidth is the major cause for the inefficiency of
the simple vector operations (a)-(c), then using faster memories
would increase the speeds of (a)-(c) more significantly than that of (d).
From Table \ref{tab:time}, we can compare the speedups of these four
vector operations as the (slower) PC800 is repalced
with (faster) PC1066. We find that the speedup for (a)-(c) is $27\%$,
but that for (d) is only $11\%$. Thus the speedups are
consistent with above picture.

Obviously, the inefficiency of vector operations (a)-(c) should
exist in any platforms, not only for the Pentium 4 systems.
To check this, we measure $ t_a, t_b, t_c $ and $ t_d $
for IBM SP2 SMP (Power 3 at 375 Mhz)  
and DEC alpha XP1000 (21264A at 667 Mhz) respectively.
The results are listed in Table \ref{tab:time_1}, which give
\bea
\label{eq:nTb}
n_T \simeq \LL\{ \begin{array}{ll}
                 21, &  \mbox{DEC alpha XP1000}, \\
                 20, &  \mbox{IBM SP2 SMP}. 
                 \end{array} \RR.
\eea

\begin{table}
\caption{
 Similar to Table \ref{tab:time}, except for the platforms:
 IBM SP2 SMP (Power 3 at 375 Mhz) with 4 Gbyte memory,
 and DEC alpha XP1000 (21264A at 667 Mhz) with 1.5 Gbyte memory.}
\begin{center}
\begin{tabular}{c|c|c|cc}
& Operation & \# of FPO & \multicolumn{2}{c}{CPU time(ns) per FPO}
  \\
& &  & IBM & DEC \\
\hline
\hline
(a) & $ |A\rangle = c_1|A\rangle + c_2 |B\rangle $
    & $ 72 \times N_{site} $
    & 5.269 & 7.232   \\
(b) & $ |V\rangle = |A\rangle + c |B\rangle $
    & $ 48 \times N_{site} $
    & 10.98 & 12.91  \\
(c) & $ \alpha = \langle V|V\rangle $
    & $ 36 \times N_{site}$
    & 6.209 & 7.684   \\
(d) & $ |A\rangle = H_w | B \rangle $
    & $ 1644 \times N_{site} $
    & 2.379 & 3.054   \\
\end{tabular}
\end{center}
\label{tab:time_1}
\end{table}

Although it is impossible to go through all platforms and measure
the values of $ t_a $, $ t_b $ and $ t_d $ individually,
it is expected that $ t_a/t_d > 1 $ and $ t_b/t_d > 1 $
(such that the inequality (\ref{eq:tdab}) is amply satisfied)
is a common feature of most systems.
In other words, we expect that the double pass is faster than
the single pass for $ n > n_T \simeq 12 - 25 $, at least for
most platforms.

Recall that in Neuberger's test run with
SGI O2000, at $ n=32 $, the double pass is faster than
the single pass by $ 30\% $ \cite{Neuberger:1998jk}.
This is not a surprise at all, in view of similar speedups
of other systems at $ n=32 $. For example, for IBM SP2 SMP or 
DEC alpha XP1000, sustituting the values of $ t_a $, $ t_b $, $ t_c $ 
and $ t_d $ (from Table \ref{tab:time_1}) into (\ref{eq:speedup}), 
we find that $ S = T_1/T_2 - 1 \simeq 30\% $
at $ n=32 $, which also agrees with the 
actual measurements given in the next section (see Table \ref{tab:P4_IBM_DEC}). 
Thus, {\it the speedup $ S $ of the double pass 
for $ n > n_T $ with $ n_T $ quite smaller than $ n_F $ 
is a generic feature of any platform, stemming from the fact that 
the vector operations in the 
second pass is more efficient than those (\ref{eq:Pl})-(\ref{eq:Zl})
in the single pass (i.e., $ t_a > t_d $ and $ t_b > t_d $).}

Nevertheless, {\it the salient feature of (\ref{eq:F2a}) and (\ref{eq:T2a})
is that the number of floating point operations and the CPU time
for the double pass are almost independent of $ n $.}
Thus one can choose $ n $ as large as one wishes, 
with only a negligible overhead.
For example, for the $ 16^3 \times 32 $ lattice, with $ L_{i} = 1000 $,
$ n_{ev} = 20 $, and $ q=0.95 $, the increment of $ T_2 $
from $ n=16 $ to $ n=200 $ is less than $ 0.05 \% $.
In other words, one can approximate $ (H_w^2)^{-1/2} Y $
(i.e., preserve the chiral symmetry) to any precision as ones wishes,
without noticeably extra costs.
{\it This is the virtue of Neuberger's 
double pass algorithm, which may have been overlooked in the last 
five years.}

\section{Tests}

In this section, we perform several tests on the single and the double
pass algorithms, and compare the theoretical 
thresholds $ n_T $ (\ref{eq:nT}) and $ n_F $ (\ref{eq:nF})
with the measured values.

In Table \ref{tab:perf_p}, we list the number of
floating point operations and the
CPU time for computing one column of the inverse of
\BAN
D(m_q) = m_q + (m_0 - m_q/2) ( 1 + \gamma_5 S(H_w) ) \ ,
\EAN
i.e., $ D^{-1}(m_q) = D(m_q)^{\dagger} Y $, where $ Y $ is solved 
from 
\bea
\label{eq:outer_CG_m}
D(m_q) D^{\dagger}(m_q) Y
= \left\{m_q^2+(m_0^2-m_q^2/4)[2+(\gamma_5 \pm 1) S(H_w)] \right\} Y=\Id
\eea
with multi-mass (outer) conjugate gradient
for a set of 16 bare quark masses ($ 0.02 \le m_q \le 0.3 $),
while the inner CG (\ref{eq:inner_CG}) is iterated with the single 
pass and the double pass respectively. The tests are performed on
the $ 8^3 \times 24 $ lattice with SU(3) gauge configuration generated
by the Wilson gauge action at $ \beta = 5.8 $.
Other parameters are: $ m_0 = 1.30 $,
$ n_{ev} = 32 $ (the number of projected eigenmodes), 
$ \lambda_{max}/\lambda_{min} = 6.207/0.198 $ (after projection),
and the tolerances for the outer and inner CG loops 
are $ 1.0 \times 10^{-11} $ and $ 2.0 \times 10^{-12} $ respectively.
The total number of iterations $ L_o $ for the outer CG loop is around
$ 100 \sim 103 $, while the average number of iterations for the inner 
CG loop is $ \sim 287 $.
 
With the formulas (\ref{eq:F1})-(\ref{eq:T2}), we can estimate 
the number of floating point operations and
the CPU time for computing
one column of $ D^{-1}(m_q) $ for a number $ n_q $ of bare quark masses.
For the number of floating point operations, our results are 
\bea
\label{eq:G}
G_k &=& (L_{o} + n_q) F_k +  \nn
&& N_{site} (60 L_o n_q + 84 L_o + 66 n_q) + 16 L_o n_q - 13 L_o + 18 n_q + 2
\eea
where $ L_o $ is the number of iterations of the outer CG loop
(\ref{eq:outer_CG_m}), the subscript $ k=1 (2) $ stands for
the single (double) pass respectively. Obviously, the most significant
part of $ G_k $ is the first term on the r.h.s. of (\ref{eq:G}), thus
\bea
\label{eq:Gi_a}
G_{k} \simeq (L_o + n_q ) F_k \ , \hspace{4mm} k = 1, 2 \ .
\eea
Similarly, the most significant part of the CPU time is 
\bea
\label{eq:Hi_a}
V_{k} \simeq (L_o + n_q ) T_k \ , \hspace{4mm} k = 1, 2 \ .
\eea
where $ T_1 $ and $ T_2 $ are given in (\ref{eq:T1})-(\ref{eq:T2}).

In Table \ref{tab:perf_p}, the estimated CPU times $ V_1 $ and $ V_2 $
are in good agreement with the measured CPU times 
(the deviation is always less than 5\%). 
By comparing the CPU times for the single pass and the double pass, 
we see that the double pass becomes faster than the 
single pass at $ n \simeq 13 $, in agreement
with the theoretical estimate (\ref{eq:nTa}) for $ p = 0.48 $,
where $ p $ is obtained by measuring the effective number 
of the ($n-1$) vector pairs $ \{ P^{(l)}, Z^{(l)}, l=2,\cdots,n \} $
which are updated before $ Z^{(l)} $ converges.

Further, comparing $ G_2 $ and $ G_1 $, we see that
$ G_1 \simeq G_2 $ at $ n_F \simeq 59 $, in agreement
with the theoretical estimate (\ref{eq:nFa}) for $ p = 0.48 $.

\begin{table}
\caption{The number of floating point operations and
 the CPU time (in unit of second) for Pentium 4 (2.53 GHz) with one
 Gbyte Rambus (PC1066) to compute one column of $ D^{-1}(m_q) $ for
 16 quark masses, versus the degree $ n $ of the rational
 polynomial $ R^{(n-1,n)} $ in polar approximation (\ref{eq:S_polar}).}
\begin{center}
\begin{tabular}{c|ccc|ccc|c}
 & \multicolumn{3}{c}{Double Pass}
 & \multicolumn{3}{c}{Single Pass}
 & \\
 & \# of FPO & \multicolumn{2}{c}{CPU Time(s)}
 & \# of FPO & \multicolumn{2}{c}{CPU Time(s)} & $\sigma$ \\
$ n $ & $G_2$ & $V_2$ & measured & $G_1$ & $ V_1 $ & measured & polar \\
\hline
\hline
 12 & $ 2.90 \times 10^{12} $  & 2456 & 2451
    & $ 1.68 \times 10^{12} $  & 2342 & 2241
    & $ 6 \times 10^{-5} $   \\
 13 & $ 2.90 \times 10^{12} $  & 2456 & 2452
    & $ 1.71 \times 10^{12} $  & 2429 & 2372
    & $ 3 \times 10^{-5} $   \\
 14 & $ 2.90 \times 10^{12} $  & 2456 & 2454
    & $ 1.75 \times 10^{12} $  & 2515 & 2520
    & $ 1 \times 10^{-5} $   \\
 16 & $ 2.90 \times 10^{12}$   & 2456 & 2454
    & $ 1.81 \times 10^{12}$   & 2689 & 2714
    & $ 3 \times 10^{-6}  $ \\
 32 & $ 2.90 \times 10^{12}$   & 2458 & 2456
    & $ 2.25 \times 10^{12}$   & 4097 & 4089
    & $ 3  \times 10^{-11} $ \\
 34 & $ 2.90 \times 10^{12}$   & 2458 & 2458
    & $ 2.30 \times 10^{12}$   & 4273 & 4278
    & $ 7  \times 10^{-12} $ \\
 40 & $ 2.90 \times 10^{12}$   & 2458 & 2456
    & $ 2.45 \times 10^{12}$   & 4803 & 4819
    & $ 1 \times 10^{-13}  $ \\
 56 & $ 2.90 \times 10^{12}$   & 2460 & 2460
    & $ 2.86 \times 10^{12}$   & 6218 & 6261
    & $ 2 \times 10^{-14}  $ \\
 59 & $ 2.90 \times 10^{12}$   & 2460 & 2460
    & $ 2.93 \times 10^{12}$   & 6483 & 6491
    & $ 2 \times 10^{-14}  $ \\
 60 & $ 2.90 \times 10^{12}$   & 2460 & 2461
    & $ 2.96 \times 10^{12}$   & 6572 & 6604
    & $ 2 \times 10^{-14}  $ \\
 64 & $ 2.90 \times 10^{12}$   & 2460 & 2461
    & $ 3.06 \times 10^{12}$   & 6926 & 6965
    & $ 2 \times 10^{-14}  $ \\
\end{tabular}
\end{center}
\label{tab:perf_p}
\end{table}

Also, in Table \ref{tab:perf_p},  
the remarkable feature of the double pass algorithm is demonstrated: 
{\it the number of floating point operations} ($G_2$)
{\it and the CPU time are almost independent of $ n $.}
Thus $ n $ can be increased to 64 or any higher value such that the
chiral symmetry is preserved to any precision as one wishes.
The chiral symmetry breaking or the error of the rational 
approximation $ R^{(n-1,n)} $ due to a finite $ n $ can be measured by
\bea
\sigma = \max_{Y} \left| \frac{W^{\dagger}W}{Y^{\dagger}Y}-1 \right|,
\hspace{4mm} W = S(H_w) Y \ , 
\eea
which is shown in the last column of Table \ref{tab:perf_p}.


To check the theoretical estimates for the threshold $ n_T $ in 
(\ref{eq:nTb}), we repeat the tests of Table \ref{tab:perf_p} for 
Pentium 4 (PC800), IBM SP2 SMP, and DEC alpha XP1000 
respectively. The results are listed in Table \ref{tab:P4_IBM_DEC}. 
Obviously, in each case, the double pass is faster than the single 
pass for $ n > 20 \sim 22 $, in good agreement with the theoretical 
estimates in (\ref{eq:nTb}). Further, at $ n=32 $, the speed of the 
double pass is faster than the single pass by $ 25\% $, $ 31\% $, 
and $ 31\% $ for these three platforms respectively, compatible with 
what Neuberger found in his test run with SGI O2000 \cite{Neuberger:1998jk}.
Note that for Pentium 4, using SSE2 code increases the speedup of the 
double pass to $ 66\% $ at $ n=32 $ (see Table \ref{tab:perf_p}), 
thus makes the double pass alogorithm even more favorable for P4 clusters.

\begin{table}
\caption{The CPU time (in unit of second) for the single and the double pass
 algorithms to compute one column of $ D^{-1}(m_q) $ for 16 quark masses,
 versus the degree $ n $ of the rational polynomial $ R^{(n-1,n)} $
 in polar approximation (\ref{eq:S_polar}).}
\begin{center}
\begin{tabular}{c|cc|cc|cc}
$n$ & \multicolumn{2}{c}{P4 PC800}
    & \multicolumn{2}{c}{IBM SP2 SMP}
    & \multicolumn{2}{c}{DEC alpha XP1000} \\
    &  2-pass & 1-pass & 2-pass & 1-pass & 2-pass & 1-pass \\
\hline
\hline
 20   & 4922 & 4627 & 7701 & 7674  & 9921 & 9868    \\
 21   & 4930 & 4794 & 7711 & 7881  & 9924 & 10197   \\
 22   & 4918 & 4940 & 7710 & 8090  & 9931 & 10531   \\
 24   & 4921 & 5166 & 7705 & 8529  & 9929 & 11125   \\
 26   & 4920 & 5433 & 7710 & 8990  & 9929 & 11599   \\
 32   & 4918 & 6167 & 7718 & 10138 & 9926 & 13043   \\
\end{tabular}
\end{center}
\label{tab:P4_IBM_DEC}
\end{table}

At this point, it may be interesting to repeat the tests of
Table \ref{tab:perf_p},  
but replacing the polar approximation (\ref{eq:S_polar}) 
with the Zolotarev optimal rational approximation,  
\bea
\label{eq:S_opt}
S_{opt}(H_w) = 
h_w \sum_{l=1}^{n} \frac{b'_l}{h_w^2 + c'_{2l-1}}
             \equiv H_w R_Z^{(n-1,n)}(H_w^2) \ ,  
\hspace{5mm}  h_w = H_w/\lambda_{min} \ ,
\eea
where
\bea
\label{eq:rz_n1n}
R_Z^{(n-1,n)}(H_w^2) = \frac{d'_0}{\lambda_{min}}
\frac{ \prod_{l=1}^{n-1} ( 1+ h_w^2/c'_{2l} ) }
     { \prod_{l=1}^{n} ( 1+ h_w^2/c'_{2l-1} ) }
= \frac{1}{\lambda_{min}} \sum_{l=1}^n \frac{b'_l}{h_w^2 + c'_{2l-1}} \ ,
\eea
and the coefficients $ d'_0 $, $ b'_l $ and $ c'_l $
are expressed in terms of Jacobian elliptic functions 
\cite{Akhiezer:1992,vandenEshof:2001hp,Chiu:2002eh}
with arguments depending only on $ n $
and $ \lambda_{max}^2 / \lambda_{min}^2 $
($ \lambda_{max} $ and $ \lambda_{min} $ are the maximum and the minimum
of the the eigenvalues of $ |H_w| $).
The results are listed in Table \ref{tab:perf_Z}.

\begin{table}
\caption{The number of floating point operations and the CPU
 time (in unit of second) for Pentium 4 (2.53 GHz) with one Gbyte
  Rambus (PC1066) to compute one column of $ D^{-1}(m_q) $ for 16 
  quark masses, versus the degree $ n $ of the Zolotarev rational 
  polynomial $ R_Z^{(n-1,n)} $.}
\begin{center}
\begin{tabular}{c|ccc|ccc|c}
 & \multicolumn{3}{c}{Double Pass}
 & \multicolumn{3}{c}{Single Pass}
 & \\
 & \# of FPO & \multicolumn{2}{c}{CPU Time(s)}
 & \# of FPO & \multicolumn{2}{c}{CPU Time(s)} & $\sigma$ \\
$ n $ & $G_2$ & $V_2$ & measured & $G_1$ & $ V_1 $ & measured & Zolotarev \\
\hline
\hline
 12 & $ 2.90 \times 10^{12}$   & 2456 & 2450
    & $ 1.72 \times 10^{12} $  & 2309 & 2274
    & $ 7 \times 10^{-11} $ \\
 13 & $ 2.90 \times 10^{12}$   & 2456 & 2452
    & $ 1.75 \times 10^{12} $  & 2398 & 2404
    & $ 8 \times 10^{-12} $ \\
 14 & $ 2.90 \times 10^{12}$   & 2456 & 2455
    & $ 1.78 \times 10^{12} $  & 2485 & 2463
    & $ 1 \times 10^{-12} $ \\
 16 & $ 2.90 \times 10^{12}$   & 2456 & 2455
    & $ 1.83 \times 10^{12}$   & 2659 & 2638
    & $ 3 \times 10^{-14} $ \\ 
 32 & $ 2.90 \times 10^{12}$   & 2458 & 2458
    & $ 2.25 \times 10^{12}$   & 4058 & 4068
    & $ 3 \times 10^{-14} $ \\ 
 34 & $ 2.90 \times 10^{12}$   & 2458 & 2458
    & $ 2.30 \times 10^{12}$   & 4233 & 4245
    & $ 3 \times 10^{-14} $ \\ 
 40 & $ 2.90 \times 10^{12}$   & 2458 & 2460
    & $ 2.45 \times 10^{12}$   & 4759 & 4795
    & $ 3 \times 10^{-14} $ \\ 
 56 & $ 2.90 \times 10^{12}$   & 2460 & 2462
    & $ 2.86 \times 10^{12}$   & 6159 & 6180
    & $ 3 \times 10^{-14} $ \\ 
 59 & $ 2.90 \times 10^{12}$   & 2460 & 2462
    & $ 2.92 \times 10^{12}$   & 6423 & 6459
    & $ 3 \times 10^{-14} $ \\ 
 60 & $ 2.90 \times 10^{12}$   & 2460 & 2460
    & $ 2.95 \times 10^{12}$   & 6510 & 6544
    & $ 3 \times 10^{-14} $ \\ 
 64 & $ 2.90 \times 10^{12}$   & 2460 & 2462
    & $ 3.05 \times 10^{12}$   & 6860 & 6903
    & $ 3 \times 10^{-14} $ \\ 
\end{tabular}
\end{center}
\label{tab:perf_Z}
\end{table}

Comparing Table \ref{tab:perf_p} with Table \ref{tab:perf_Z}, 
it is clear that for the single pass with $ n < 32 $, 
Zolotarev optimal approximation is better than the polar approximation, 
in terms of the precision of the approximation ($\sigma$).
However, for the double pass, the polar approximation seems to be   
as good as the Zolotarev approximation since the degree $ n $
can be pushed to a very large value, with negligible
extra CPU time. In other words, {\it with the double pass algorithm,
it does not matter which rational approximation
one uses to compute $ D^{-1} (m_q) $ in a gauge background.}
This seems to be a rather unexpected result.

\section{Concluding remarks}

So far, we have restricted our discussions to the sign function with 
argument $ H_w $. However, it is clear that the salient features of 
the double pass algorithm are invariant for other choices of the 
argument, e.g., improved Wilson operator. In general, the double pass
algorithm is a powerful scheme for the matrix-vector product 
$ R(H^2) \cdot Y $, where $ R $ can be any rational polynomial $ R $ 
with argument $ H^2 $ (positive definite Hermitian operator),  
not just for $ (H^2)^{-1/2} $.

The virtue of Neuberger's double pass algorithm is its constancy in  
speed and memory storage for any degree $ n $ of the rational
approximation, where its constancy in speed is valid under a mild 
condition ($ N_{site} \gg L_i $) which can be fulfilled in most cases. 
Further, the double pass is faster than the single pass even for $ n $
as small as 12 (Pentium 4), and it is expected that the threshold 
$ n_T \simeq 12 - 25 $ for most systems. 
Thus, it seems that there is not much room 
left for the single pass algorithm with Zolotarev approximation, unless 
the number of inner CG iterations is exceptionally large, which could 
happen if the low-lying eigenmodes of $ H_w^2 $ are not projected out 
and treated exactly.

Note that $ H_w^2 $ can be tridiagonalized by the conjugate gradient
(\ref{eq:A})-(\ref{eq:P1}), with the unitary transformation matrix 
$ U $ formed by the normalized residue vectors 
$ \{ {\hat R}_j, j=0,\cdots,i \} $, and the elements of the 
tridiagonal matrix expressed in terms of the coefficients 
$ \{\alpha_j, \beta_j, j=0,\cdots, i \} $ \cite{Golub:83} 
(up to the tolerance of the conjugate gradient), i.e.,   
\bea
{U}^{\dagger} H_w^2 {U} \simeq {\cal T}
\eea
where 
\bea
{U}_{kj} = \frac{(R_j)_k}{\sqrt{\LL< R_j | R_j \RR>}} \ , 
\eea
and $ {\cal T} $ is a symmetric tridiagonal matrix with nonzero elements 
\bea
{\cal T}_{jj} &=& \frac{\beta_j}{\alpha_{j-1}} + \frac{1}{\alpha_j} \ , \\
{\cal T}_{j+1,j}={\cal T}_{j,j+1} &=& -\frac{\sqrt{\beta_{j+1}}}{\alpha_j} \ ,
\hspace{4mm} j=0,\cdots,i.
\eea
Thus, after running the first pass of the CG loop 
(\ref{eq:A})-(\ref{eq:P1}), $ {\cal T} $ can be constructed  
from the coefficients $ \{ \alpha_j, \beta_j \} $, and  
diagonalized by an orthogonal transformation 
\bea
{\cal T} = O \Lambda \tilde{O} \ .
\eea
Then the matrix-vector product $ (H_w^2)^{-1/2} Y $ can be 
evaluated as     
\bea
\label{eq:cgs}
\frac{1}{\sqrt{H_w^2}} Y \simeq 
{U} O \frac{1}{\sqrt{\Lambda}} \tilde{O} {U}^{\dagger} Y   
= \sum_{j=0}^i l_j R_{j} 
\eea
where 
\bea
\label{eq:lj}
l_j = \sum_{m=0}^i 
      O_{jm} \frac{1}{\sqrt{\lambda_m}} O_{0m} \ 
      \sqrt{\frac{\LL< R_0|R_0 \RR>}{\LL< R_j|R_j \RR>}} \ .
\eea
Here the summation on the r.h.s. of (\ref{eq:cgs}) is obtained  
by running the second pass of the CG loop 
$ \{ (\ref{eq:A}),(\ref{eq:B}),(\ref{eq:R}),(\ref{eq:P1}) \} $,
and adding $ l_j R_j $ to the sum successively from $j=0$ to $i$.

It is well known that (any positive definite Hermitian matrix) 
$ H_w^2 $ can be tridiagonalized by
Lanczos iteration \cite{Golub:83,Cullum:85} as well as the 
conjugate gradient. The connection 
between the Lanczos iteration and the conjugate gradient for the  
tridiagonalization of a positive definite Hermitian matrix has 
been well established \cite{Golub:83}, and both have almost the   
same performance in practice. 
In Ref. \cite{Borici:1999ws}, the Lanczos approach was proposed
for the matrix-vector product $ (H_w^2)^{-1/2} Y $, and its variant 
(replacing Lanczos iteration with the conjugate gradient)
was used in Ref. \cite{Gavai:2001vx}. 

The only difference between the Lanczos (CG) algorithm and
Neuberger's double pass algorithm is the diagonalization of
the tridiagonal matrix $ {\cal T} $ and the computation 
of the coefficients $ \{ l_j \} $ (\ref{eq:lj}) in the former
versus the computation of the coefficients $ \{ c_j \} $ 
(\ref{eq:cij}) in the latter. Since the number of floating point
operations for the diagonalization of a symmetric tridiagonal 
matrix $ {\cal T} $
is $ \simeq 3 L_i^3 $ (where $ L_i $ is the number of iterations
of the inner CG loop, or the size of $ {\cal T} $), it is compatible 
with that of computing the coefficients $ \{ c_j \} $, i.e., 
the last term on the r.h.s. of (\ref{eq:F2}). Thus we expect that
the performance (speed and accuracy) of Lanczos (CG) algorithm and 
Neuberger's double pass algorithm are compatible.

In Table \ref{tab:lanczos}, we compare the Lanczos (CG)
algorithm with Neuberger's double pass algorithm, by computing one column 
of $ D^{-1}(m_q) $ (for 16 bare quark masses) on the $ 16^3 \times 32 $ 
lattice with SU(3) gauge configuration generated by the Wilson gauge action 
at $ \beta = 6.0 $. Other parameters are: $ m_0 = 1.30 $,
$ n_{ev} = 20 $ (the number of projected eigenmodes), 
$ \lambda_{max}/\lambda_{min} = 6.260/0.215 $ (after projection),
and the tolerances for the outer and inner CG (Lanczos) loops 
are $ 1.0 \times 10^{-11} $ and $ 2.0 \times 10^{-12} $ respectively.
The number of iterations for the outer CG loop is $ L_o = 347 $, 
while the average number of iterations for the inner 
CG loop is $ \sim 300 $.
Evidently, these seemingly different algorithms have almost 
the same speed as well as the accuracy ($\sigma$).

\begin{table}
\caption{The number of floating point operations and the CPU
 time (in unit of second) for Pentium 4 (2.53 GHz) with one Gbyte
  Rambus (PC1066) to compute one column of $ D^{-1}(m_q) $, versus 
  different algorithms.} 
\begin{center}
\begin{tabular}{c|cc|cc}
 & \multicolumn{2}{c}{Double pass algorithm}
 & \multicolumn{2}{c}{Lanczos (CG) algorithm}
  \\
 & Polar($n=128$) & Zolotarev($n=16$) & Lanczos & CG  
    \\
\hline
\hline
FPO & $9.49 \times 10^{13}$ & $9.49 \times 10^{13}$    
    & $9.54 \times 10^{13}$ & $9.51 \times 10^{13}$ \\  
Time (total)    & 94543 & 94632 & 97824 & 94722 \\ 
Time (2nd pass) & 46281 & 46303 & 46353 & 46174 \\
$ \sigma $ & $ 1 \times 10^{-14} $ 
           & $ 1 \times 10^{-14} $ 
           & $ 1 \times 10^{-14} $ 
           & $ 1 \times 10^{-14} $ \\
\hline
\end{tabular}
\end{center}
\label{tab:lanczos}
\end{table}

Thus, for quenched lattice QCD, one has several compatible options
to compute the quenched quark propagator
\bea
(D_c + m_q )^{-1} = ( 1 - r m_q )^{-1} [ D^{-1}(m_q) - r ] \ ,
\hspace{4mm} r = \frac{1}{2m_0} \ , 
\eea 
even though we have chosen Neuberger's double pass algorithm to solve
$ D^{-1}(m_q) $ in our recent investigation \cite{Chiu:2003iw}.
Nevertheless, for lattice QCD with dynamical quarks,
the quark determinant $ \det D(m_q) $ could not be computed directly 
with existing algorithms and computers. 
If $ \det D(m_q) $ is incorporated through the dynamics of $ 2n $ 
pseudofermion fields
(where $ n $ can be regarded as the degree $ n $ in the rational 
polynomial $ R^{(n-1,n)} $), then an additional degree of
freedom (or the fifth dimension with $ N_s = 2n $ sites) 
has to be introduced. Thus a relevant question is how to reproduce  
$ \det D(m_q) $ accurately with the minimal $ N_s $.
A solution has been presented in Ref. \cite{Chiu:2002ir}.
On the other hand, it would be interesting to see whether there is 
an algorithm to drive the dynamics of these $ N_s $ pseudofermion fields
such that the cost is almost independent of $ N_s = 2n $.

\bigskip
\bigskip
\flushpar
{\bf Acknowledgement }
\bigskip

\noindent

T.W.C. would like to thank Herbert Neuberger and Rajamani Narayanan
for a brief but stimulating discussion at the 
Institute for Advanced Study, during a visit in November 2002.  
This work was supported in part by the National Science Council,
ROC, under the grant number NSC91-2112-M002-025.


\vfill\eject

\end{document}